# Where are the Dangerous Intersections for Pedestrians and Cyclists: A Colocation-Based Approach[1]

Yujie Hu[a*], Yu Zhang[b], Kyle S. Shelton[c]

[a] School of Geosciences, University of South Florida, Tampa, FL 33620, USA
[b] Department of Civil and Environmental Engineering, University of South Florida, Tampa, FL 33620, USA
[c] Kinder Institute for Urban Research, Rice University, Houston, TX 77005, USA
*Corresponding author: Yujie Hu

**Abstract**
Pedestrians and cyclists are vulnerable road users. They are at greater risk for being killed in a crash than other road users. The percentage of fatal crashes that involve a pedestrian or cyclist is higher than the overall percentage of total trips taken by both modes. Because of this risk, finding ways to minimize problematic street environments is critical. Understanding traffic safety spatial patterns and identifying dangerous locations with significantly high crash risks for pedestrians and cyclists is essential in order to design possible countermeasures to improve road safety. This research develops two indicators for examining spatial correlation patterns between elements of the built environment (intersections) and crashes (pedestrian- or cyclist-involved). The global colocation quotient detects the overall connection in an area while the local colocation quotient identifies the locations of high-risk intersections. To illustrate our approach, we applied the methods to inspect the colocation patterns between pedestrian- or cyclist-vehicle crashes and intersections in Houston, Texas and we identified among many intersections the ones that significantly attract crashes. We also scrutinized those intersections, discussed possible attributes leading to high colocation of crashes and proposed corresponding countermeasures.

## 1. Introduction
Active transportation modes such as walking and bicycling are critical elements of a sustainable urban transportation system. Many U.S. cities, including Houston, are

---

grappling with how to reduce vehicular traffic, energy consumption, and greenhouse gas emissions that come with car travel. Beyond their clear environmental benefits, the large-scale embrace of active modes of travel could lead to significant gains in public health (Krizec 2007; Miranda-Moreno et al. 2011).

Automobile-centric street design of most U.S. cities, however, leaves pedestrians and cyclists vulnerable. For example, pedestrians were reported to be 23 times more likely to be killed than vehicle occupants in terms of distance traveled (Pucher and Dijkstra 2003; Moudon et al. 2008). According to the National Highway Traffic Safety Administration (2015), an average of 4,500 pedestrians were killed and 66,000 injured in automobile-related crashes each year between 2004 and 2013. Nationally, pedestrians are involved in 14% of all fatality crashes, despite the fact that walking remains a minimal part of the overall share of trips taken (National Highway Traffic Safety Administration 2012a). As for cyclists, the percentage of cyclist fatalities increased from 1.5% in 2003 to 2.2% in 2012 (National Highway Traffic Safety Administration 2012b).

The majority of pedestrian/cyclist crash injuries and fatalities occurred in cities, where activity is concentrated and where cycling and walking rates are at their highest (National Highway Traffic Safety Administration 2008). Due in part to high transportation costs, re-urbanization is occurring in metro regions across the U.S. and more people are moving back to cities. As urban populations grow and the streetscape maintains the same car-centric design, these vulnerable road users will still likely face high crash risks.

Some studies have employed density analysis to look for crash clustering patterns and to identify the most dangerous locations (e.g., intersections) based on historical crash incidents. One such example is to plot the histogram of absolute value of crashes or crash rates and take the locations with statistically high values as the most dangerous locations (e.g., Pulugurtha et al. 2007; Anderson 2009; Ferreira and Couto 2015). Another set of studies take that approach a step further by focusing on understanding the connection between elements of the built environment (e.g., intersections) and crashes by using macro-level regression analysis (e.g., Moudon et al. 2008; Ukkusuri et al. 2012). Most of these studies pre-define the connection between crashes and intersections by assigning crashes to intersections and take one intersection as one data point in the regression analysis. Both these methods have limitations that, if used to inform policy or funding, may lead to ineffective safety countermeasures being implemented.

Using a technique called colocation pattern mining within Geographic Information Systems (GIS), this paper examines the spatial correlation pattern between crashes and intersections to offer a novel way to understand at which intersections crashes are more likely to occur. To illustrate our approach, two colocation measures, global and local, are used to analyze the spatial relationship between crash incidents and intersections in the city of Houston based on the pedestrian- and cyclist-vehicle crash records between 2010 and 2016. Our analysis provides an alternative way to understand the spatial links between crash incidents and intersections. In particular, our local colocation method is able to present stakeholders with site-specific patterns and identify the most dangerous locations.

## 2. Literature Review

Although traffic crashes have been widely studied by the transportation research community, crashes that involve automobiles and either pedestrians or cyclists are still understudied (Ukkusuri et al. 2012). Most existing research in this area has relied on clustering analysis. For example, Pulugurtha et al. (2007) employed the kernel density estimation (KDE) technique to detect pedestrian crash zones and Anderson (2009) discovered and classified crash hotspots for pedestrians, cyclists, and vehicles, based on KDE and K-means clustering approach. However, Dai and Jaworski (2016) criticized the application of the above planar KDE in analyzing traffic accidents that occur in a linear space like a road, and instead offered a network-based KDE method to locate pedestrian crash hotspots. Other clustering methods used to detect pedestrian or cyclist crash hotspots in past research include the global and local spatial autocorrelation (e.g., Truong and Somenahalli 2011), the spatiotemporal Bernoulli model (Dai 2012), and the binary Probit model (Ferreira and Couto 2015).

Another line of crash research shifted the focus from locating crash hotspots to studying the links between the urban built environment and crashes. Commonly examined built environment elements include urban land use type or mix, road network characteristics, and accessibility to public transit systems and educational facilities. Typical techniques for analyzing the links between crashes and the built environment are regression models such as the negative binomial model, the log-linear model, the Poisson model and the Bayesian model (e.g., Li et al. 2007; Moudon et al. 2008; Pulugurtha and Sambhara 2011; Ukkusuri et al. 2012; Chen and Zhou 2016). For example, Loukaitou-Sideris et al. (2007) reported that areas proximal to educational facilities and with higher fractions of commercial land use are correlated with more pedestrian crashes. Similarly, Ukkusuri et al. (2012) found that census tracts with denser commercial or industrial land use types had more crashes than those with a greater portion of residential land use. In addition, they discovered that road characteristics such as the width and the number of lanes are other related factors for pedestrian crashes. Chen (2015) revealed that traffic analysis zones (TAZ) with a greater number of road signals and street parking signs were likely to see more bicycle crashes. Further, Chen and Zhou (2016) reported that higher densities of four-way and signalized intersections, mixed land use, and more bus stops in a TAZ would result in a higher pedestrian crash frequency; in contrast, the density of sidewalk was negatively related to the crash frequency. Besides purely built environment elements, some studies also considered interactions between built environment variables and socio-demographic characteristics such as proportions of different age groups (e.g., children and senior population) or race-ethnicity groups like the black and Hispanic population (Cho et al. 2009; Ukkusuri et al. 2012; Chen 2015).

Despite the increasing attention given to studying the impacts of the built environment on crashes, most of these efforts are still focused on the macro level–TAZs, census tracts or even counties (e.g., Loukaitou-Sideris et al. 2007; Wier et al. 2009; Huang et al. 2010; Ukkusuri et al. 2012; Narayanamoorthy et al. 2013; Chen 2015; Chen and Zhou 2016). Analyses at the micro level such as individual intersections can provide more detailed and localized patterns than area-wide analyses. Such close looks can help decision-makers propose site-specific strategies for safety improvements (Son et al. 2011; Ukkusuri et al. 2012). In addition, analyses at a disaggregated level can lead to more consistent and reliable results due to the well-known modifiable areal unit problem (MAUP) in geospatial studies, which would cause inconsistent findings in quantitative measures and statistical tests when analyzing problems at different scales such as TAZs and tracts (e.g., Hu and Wang 2015; 2016). A few studies have attempted to analyze the impacts of the built environment on the crash frequency at the micro level using statistical approaches (e.g., Lee and Abdel-Aty 2005; Lee et al. 2017; Murphy et al. 2017); however, they only focused on a sample of signalized intersections. Additionally, the statistical approaches—which model the crash frequency by taking crash site attributes as explanatory variables—are often used to infer causation from the association and to predict future trends by controlling for certain attributes. Lord and Mannering (2010) reviewed more than one hundred papers that applied statistical approaches to model crash-frequency data and discussed the strengths and limitations of different statistical models (refer to Table 2 on page 295 of Lord and Mannering 2010). They concluded that although crash-frequency data analysis has been advanced over the years, the work was still inherently limited by the available data. To impel the microscopic crash-frequency analysis, detailed driving data and crash data from a vehicle's black box would be needed.

Colocation measures the spatial correlation between point objects. Different from the spatial autocorrelation that examines the association between objects of the same type, colocation targets objects of different types such as crime vs. land use facility (Wang et al. 2017). In a word, it measures if one type of objects is spatially attracted to another type. Colocation pattern analysis is commonly used in business, crime, and ecology studies. For example, Leslie and Kronenfeld (2011) examined if business establishments colocate near establishments of the same economic sector or different sectors. The authors also applied the method to investigate the colocation pattern between types of trees in terms of different health conditions (e.g., multistem, leaning, and dead). Cromley et al. (2014) analyzed the colocation pattern of different types of housing (e.g., colonial, cape, and ranch). Wang et al. (2017) investigated the colocation pattern of land use facilities and different types of crime, i.e., what types of land use facilities (e.g., school, retail shop, and entertainment establishment) significantly attracted what types of crime (e.g., residential burglary, robbery, and motorcycle theft).

The techniques to measure colocation patterns can be grouped into global and local measures. Popular global measures include the cross K (or bivariate K) function. This approach compares the difference between the observed overall density of type B objects within a predetermined distance of a type A object with what we would expect under the assumption that the two object populations are independent (Cuthbert and Anderson 2002). The independence assumption is critical. If it is not met, the cross K function will return highly biased results; for example, it is very likely to find a significant colocation pattern between two types of objects when they collectively cluster together. Therefore, results from this type of methods would be highly unreliable. Recently, Leslie and Kronenfeld (2011) designed the colocation quotient (CLQ), which examines the overall association between the observed number of type B objects in proximity to type A objects and the expected number by chance. Different from the cross K function, the CLQ uses the topological distance (i.e., nearest neighbors) rather than the actual metric distance (e.g., Euclidean distance) to account for the effect of the joint population clustering layout (Wang et al. 2017). The CLQ is by design a global measure and thus termed the global CLQ (GCLQ). Later, Cromley et al. (2014) extended the GCLQ to a local version (LCLQ) that takes into account the spatial heterogeneity of the association between types of objects under investigation. Wang et al. (2017) further improved it by proposing a statistical significance test to attach statistical confidence to the derived LCLQ.

This article studies how the built environment affects crashes between automobiles and either a pedestrian or cyclist from a spatial correlation perspective using two colocation indicators, the GCLQ and LCLQ. Our research, to the best knowledge of authors, is the first effort that extends the colocation analysis into the transportation safety field and examines the spatial correlation patterns between crashes and intersections (both signalized and non-signalized) at the micro level. The merits of our study are as follows. First, our method demonstrates the viability of studying spatial correlation patterns to identify specific locations of dangerous intersections. This is an alternative approach from existing micro-level regression models. Second, compared to area-level methods prevalent in the existing literature, our model can provide site-specific patterns and it is not affected by the aforementioned MAUP effect. Third and finally, our analysis is targeted at a full set of intersections, instead of only signalized intersections.

## 3. Methodology
### 3.1 Global Colocation Quotient

Based on the well-known economic statistics location quotient (Stimson et al. 2006), Leslie and Kronenfeld (2011) designed the GCLQ to measure the overall colocation pattern between any two types of point objects—in our case recorded crashes and intersections. To search for colocation, the analysis starts with the observed number of times that type B objects, say traffic light controlled-intersections, have type A objects, say fatality crashes, in close proximity to them. That number is then compared with those expected to occur by chance through a series of simulation trials.

The GCLQ is formulated as (Wang et al. 2017)

$$GCLQ_{A \to B} = \frac{N_{A \to B}/N_A}{N_B/(N-1)} \quad (1)$$

where $N_A$ is the number of type A objects (e.g., fatality crashes), $N_B$ represents the number of type B objects (e.g., traffic light controlled-intersections), $N_{A \to B}$ depicts the number of type A objects with type B objects as their nearest neighbors, and N denotes the total number of objects under investigation. The numerator estimates the observed proportion of type B objects that are the nearest neighbors of type A objects, and the denominator computes the expected proportion by chance (N – 1 is used rather than N because an object cannot be its own nearest neighbor).

Occasionally, an object A may have multiple nearest neighbors. As formulated in Equation 2, the GCLQ assigns equivalent weight to each of its nearest neighbors when calculating $N_{A \to B}$. In the formula, i represents each type A object, $nn_i$ depicts the number of object i's nearest neighbors, j denotes each of the $nn_i$ nearest neighbors, and $f_{ij}$ is a binary variable indicating whether object i's nearest neighbor j is of type B or not (1 means yes and 0 otherwise).

$$N_{A \to B} = \sum_{i=1}^{N_A} \sum_{j=1}^{nn_i} \frac{f_{ij}}{nn_i} \quad (2)$$

Given the values of N, $N_A$, and $N_B$, Equation 3 calculates the expected count of type A objects that have type B objects as their nearest neighbors by randomly allocating the categories of all objects under investigation. Specifically, it equals $N_A$ multiplied by the conditional probability of selecting a type B object given a type A object already selected. Substitute the expected value of $N_{A \to B}$ in Equation 3 into Equation 1, we then have Equation 4 that calculates the expected value of $GCLQ_{A \to B}$ (Leslie and Kronenfeld 2011).

$$E(N_{A \to B}) = N_A E(P_{B|A}) = \frac{N_A N_B}{N-1} \quad (3)$$

$$E(GCLQ_{A \to B}) = E\left(\frac{N_{A \to B}/N_A}{N_B/(N-1)}\right) = E(N_{A \to B})\frac{N-1}{N_A N_B} = \left(\frac{N_A N_B}{N-1}\right)\left(\frac{N-1}{N_A N_B}\right) = 1$$

(4)

Clearly, $GCLQ_{A \to B}$ has an expected value of one when all objects are randomly relabeled given the spatial and frequency distributions of each object population. Therefore, a $GCLQ_{A \to B}$ greater than one indicates a possible colocation pattern (e.g., fatalities tend to colocate with traffic light controlled-intersections), and a larger GCLQ value means stronger colocation. On the contrary, a $GCLQ_{A \to B}$ less than one denotes a possible isolation pattern (e.g., fatalities tend to be separated from traffic light controlled-intersections), and a less GCLQ score implies greater isolation pattern.

This analysis initially only suggests a possible colocation (or isolation) pattern between two types of objects without statistical significance. Then a Monte Carlo simulation technique is used to compare the observed GCLQ with the null hypothesis, which states that no spatial association exists between type A and type B objects given their joint distribution pattern (e.g., clustered or dispersed in general). Specifically, the simulation process randomly reassigns the category of each object by following the frequency distribution of each object population (e.g., the percentage of type A objects in the entire population remains unchanged after the relabel process). By repeating this process many times such as 1,000 runs, we will have a sample distribution of the $GCLQ_{A \to B}$. Finally, we compare the distribution of the observed $GCLQ_{A \to B}$ with the corresponding sample distribution to obtain a test statistic along with a significance level (Leslie and Kronenfeld 2011; Kronenfeld and Leslie 2015; Wang et al. 2017).

When searching for the nearest neighbors of each type A object (e.g., crash location), instead of the built-in Euclidean distance in the original GCLQ, we used the street network distance, given that traffic collisions were geocoded to the existing street network, and road density is highly uneven across areas (e.g., sparse in rural areas and compact in urban centers). The GCLQ values between both distance measures would be comparable in urban areas but vary substantially in rural areas (Wang et al. 2017).

### 3.2 Local Colocation Quotient

The global measure helps identify the general colocation patterns across an entire area, but the patterns may not apply to every single place in the area due to shifting elements such as population and land use. Accordingly, Cromley et al. (2014) designed a local colocation quotient (LCLQ) to examine localized patterns and Wang et al. (2017) further improved it by adding a rigorous statistical test for the confidence of results. The local analysis examines the relationship between a specific type A object and a subset of type B objects nearby. Using the same fatality and traffic light controlled-intersection examples as above, it shows how a specific fatality crash either colocates with or isolates from those intersections within a predefined catchment area of it. Therefore, the LCLQ can capture the variability of the global spatial (in)dependency across places and help better understand the spatial process and target underlying driving factors. Another advantage over the GCLQ is that the outputs of the LCLQ are maps that are more readable and understandable (Wang et al. 2017).

The LCLQ is formulated as

$$LCLQ_{A_i \to B} = \frac{N_{A_i \to B}}{N_B/(N-1)} \quad (5)$$

$$N_{A_i \to B} = \sum_{j=1(j \neq i)}^{N} \left( w_{ij} f_{ij} \middle/ \sum_{j=1(j \neq i)}^{N} w_{ij} \right) \quad (6)$$

$$w_{ij} = \exp\left(-0.5 * \frac{d_{ij}^2}{d_{ib}^2}\right) \quad (7)$$

where $A_i$ denotes the i-th A object, $f_{ij}$ indicates whether object $A_i$'s nearest neighbor(s) object j is a marked B object (=1) or not (=0), $w_{ij}$ is the weight of object j, representing the importance of object j to object $A_i$ based on two distances—$d_{ij}$ for the distance between objects $A_i$ and j, and $d_{ib}$ for the bandwidth distance to search for neighbors around object $A_i$. The other notations are the same as described above in the GCLQ. Equation 5 defines the formulation of LCLQ. Similar to the GCLQ in Equation 1, the numerator calculates the observed proportion of type B objects that are the nearest neighbors of each object $A_i$, and the denominator computes the proportion of type B objects that could be the nearest neighbors to type A objects. Equation 6 signifies the weighted average number of type B objects that are the nearest neighbors of object $A_i$, and Equation 7 denotes the Gaussian kernel density weighting function—generally, a neighbor located closer to object $A_i$ is assigned a greater weight (Cromley et al. 2014; Wang et al. 2017).

Like the GCLQ, the LCLQ uses distance ranks (i.e., topological distance) instead of actual metric distance to search for neighbors as well. This type of adaptive bandwidth setting ensures exactly the predefined number of points to be included in the LCLQ calculation at each marked A object and thus reports more reliable results than the fixed bandwidth setting based on metric distance. This is particularly beneficial to analyses targeting urban areas with varying density of intersections—street networks in general—see Figure 1 for the Houston example. Furthermore, the adoption of network distance instead of straight-line distance in ranking neighbors also contributes to improving the accuracy of results since traffic accidents were geocoded to street network.

Monte Carlo simulation is also used to test if the derived LCLQ is statistically significant from random (Wang et al. 2017). At each given object $A_i$, one simulation trial randomly relabels the category of any other objects by following the frequency distribution of each category (Kronenfeld and Leslie 2015). Take $LCLQ_{F_1 \rightarrow TL}$ as an example ($F_1$ means the first record of fatality crashes and TL denotes traffic light controlled-intersections), each simulation run randomly reassigns the labels of all objects (including fatality crashes and all types of intersections including traffic light controlled-, stop sign controlled- and noncontrolled-intersections) but not object $F_1$, and the number of objects in each category remains the same after the process. By iterating the above simulation process for multiple times (e.g., 1,000 times in our analysis), we can obtain a sample distribution of LCLQ for each given crash incident under investigation and then compare with the observed LCLQ to derive the significance level. One may refer to Leslie and Kronenfeld (2011) and Wang et al. (2017) for more technical details.

## 4. Study Area and Data Sources

The study area of our analysis is the city of Houston, Texas, the fourth-most populous city and the fourth-most dangerous city for road violence in the United States. Houston has been facing high traffic-related deaths and serious injuries. In 2014, 667 people died on streets in the Houston region with 227 of those fatalities occurring within the city of Houston. According to the Texas Department of Transportation, the percentage of pedestrian-involved crashes increased by 88% between 2010 and 2015, and cyclist-involved crashes increased by 42% during the same time period. This disturbing increase, along with other factors such as a desire to encourage active transportation, has led concerned citizens and Houston area officials to pursue actions to make existing streets safer. The city recently approved a new long-range bike plan and the previous mayor signed a complete streets executive order that the current administration has left in force.

The major data sources of this study include traffic crashes, street networks, and locations of stop signs and traffic lights. Specifically, traffic crash records were collected from Texas Department of Transportation for a six-year period between January 2010 and September 2016. The records include the latitude/longitude of the crash site, the time when the crash occurred, the injury severity (e.g., fatality, injury, and no injury), the actors involved (e.g., driver, passenger, cyclist, and pedestrian), and whether the incident occurred near an intersection. Over this time period, there were 398,813 crashes recorded in the city of Houston. Of those incidents, 390,713 (98.0%) were vehicle-vehicle crashes, 5,711 (1.4%) were pedestrian-vehicle crashes and 2,389 (0.6%) were cyclist-vehicle crashes. We extracted and geocoded the 8,100 pedestrian- or cyclist-vehicle crashes (2.0%) for subsequent analyses. During the data cleaning process, we identified some reporting errors such as when pedestrian- or cyclist-vehicle accidents that occurred on highways were reported as intersection-related accidents. Such reporting errors were removed from the dataset. From the 8,100 pedestrian- or cyclist-vehicle crashes, we finally pulled out 3,952 intersection-related records, of which, 90 resulted in fatalities, 1,802 resulted in injuries and 2,060 no injury. Figure 1A displays the locations of pedestrian- or cyclist-vehicle incidents within the city.

Street network data was obtained from Esri Inc. (Environmental Systems Research Institute). We built a network dataset in Esri ArcGIS 10.4 and created 319,793 junctions and 396,125 road segments. Furthermore, we removed the generated junctions that are not real intersections, including dangles (i.e., end points of dead-end streets), line vertices (i.e., points connecting two street segments with the same name) and ramp junctions (i.e., points where ramps merge with highways). After the cleaning up, 78,372 junctions (24.5% of the original junctions) were retained as intersections in Houston. Locations of stop signs and traffic lights were gathered from the Houston Public Works and Engineering Department. There are currently 18,882 stop signs and 2,286 traffic lights in service. It should be noted that there is only one point recorded to represent stop signs or traffic lights at an intersection so it is not possible to differentiate between two-way and four-way stop sign intersections. We performed a spatial join to relate stop signs/traffic lights to intersections; for example, an intersection is defined as a traffic light controlled-intersection if there is a traffic light within thirty feet. Eventually, we have three types of intersections—2,286 traffic light controlled-intersections, 18,882 stop sign controlled-intersections and 57,204 non-controlled intersections. Figure 1B displays the locations of these intersections.

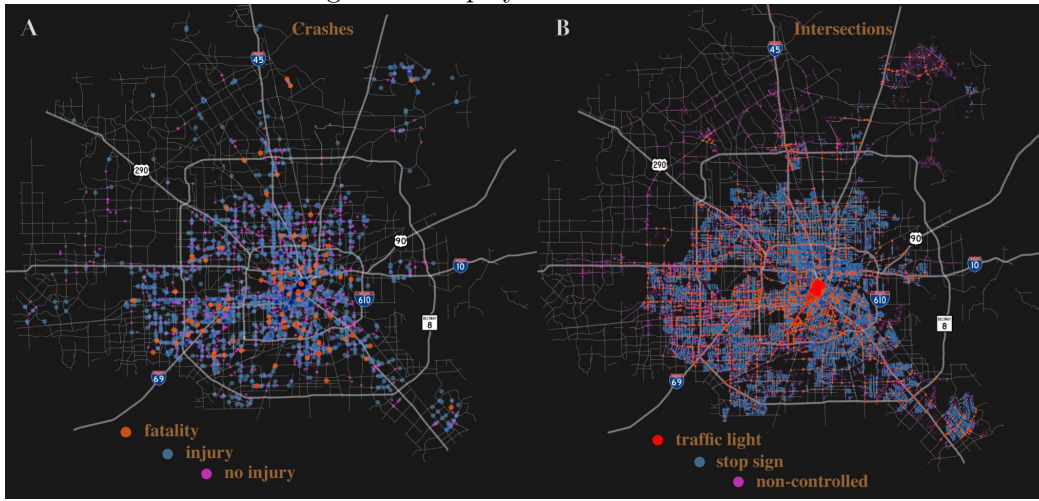

Figure 1. Locations of (A) pedestrian- or cyclist-vehicle crashes and (B) intersections in Houston

We first applied GCLQ to analyze the general spatial correlation between three types of crashes (i.e., fatality, injury, and no injury) and three types of intersections

(i.e., traffic light controlled-, stop sign controlled-, and noncontrolled-intersections). The local version, LCLQ, was then followed to provide site-specific analysis and detect the (in)stability of the general spatial correlation patterns across individual intersections.

For simplicity, pedestrian- or cyclist-vehicle crashes are referred to as crashes, and traffic light controlled-, stop sign controlled-, and noncontrolled-intersections as TL, SS, and NC intersections hereafter.

## 5. Results
## 5.1 Global Colocation Analysis

Table 1. Global colocation analysis of crashes and intersections

| Crash type | Signals | | |
| --- | --- | --- | --- |
| | NC | SS | TL |
| Fatality | 0.5*** | 1.48* | 9.81*** |
| Injury | 0.57*** | 1.49*** | 7.79*** |
| No injury | 0.58*** | 1.44*** | 7.90*** |

Note: *** significant at 0.001 level, * significant at 0.05 level.

The global colocation results are shown in Table 1. All outcomes are statistically significant at the 0.001 level, except the outcome for the fatality at SS intersection. Clearly, all three types of crashes had colocation relationship with controlled intersections. The colocation from any of the crash categories to TL intersections is remarkably greater than that to SS intersections (e.g., average GCLQs of 8.50 vs. 1.47). And the most significant pattern was between fatality crashes and TL intersections. This is consistent with the finding by Sze and Wong (2007) that crashes occurring at a signalized intersection would result in a higher risk of fatality and severe injury.

In that vein, it is also worth highlighting that any of the crash categories were highly separated from NC intersections on the global level (average GCLQ across three crash types is 0.55, far less than 1). Ossenbruggen et al. (2001) reported a similar pattern that crashes are about two times more likely to occur at an intersection with signal control than a non-controlled site. The pattern is likewise understandable because NC intersections are often located in neighborhoods where vehicle speeds are lower and pedestrians/cyclists are more visible to drivers. This suggests that enhancing the visibility of pedestrians/cyclists, increasing the awareness of drivers to the possibility of pedestrians, or changing the structure of TL intersections to encourage slower speeds may help reduce the high crash risks for them at signalized intersections.

The above global colocation analysis examines the overall colocation patterns between all crashes of a certain type and all intersections of a certain kind in the study area. It is invalid to apply the area-wide patterns to individual intersections; therefore, the global colocation analysis can only provide limited insights on countermeasures of improving traffic safety. More detailed analysis, especially at the site-specific level, is needed to understand spatial variations of the overall colocation patterns at particular intersections.

**5.2 Local Colocation Analysis**

We then applied the LCLQ to examine localized colocation patterns of incidents and intersections. Statistical significance was derived by running simulation for 1,000 times. In Figure 2, orange circles represent crash events that were significantly colocated with a certain type of intersections (at the 0.05 level), blue circles denote crash incidents that were significantly isolated from a certain type of intersections (at the 0.05 level) and pink circles present those crashes without statistical significance (for better visualization, we only show insignificant LCLQs of fatality accidents in Figures 2A, 2D and 2G as an example). Given the focus on identifying locations of high crash risks in this article, we only emphasize those orange incidents that were significantly colocated with intersections.

Between crashes and TL intersections, the global analysis found a remarkably strong colocation relationship (GCLQ = 9.81, significant at the 0.001 level). As shown in Figure 2A, the local analysis identified only two fatality incidents (mostly overlapped) in downtown Houston (intersections of Rusk Street & Louisiana Street and Rusk Street & Travis Street) that were significantly colocated with TL intersections. Remarkably, the average LCLQ of the two fatality crashes is 34.32, indicating an extremely high degree of colocation relationship with TL intersections.

The injury crashes and TL intersections exhibited an overall intense colocation pattern (GCLQ = 7.79, significant at the 0.001 level) but to a lesser degree than fatality crashes (GCLQ = 9.81). Looking at the localized patterns between injury incidents and TL intersections in Figure 2B, there were 48 injury incidents significantly attracted to TL intersections. Of these, 45 incidents were concentrated in downtown Houston and the other three were located at the intersections of Hardy Street & Cavalcade Street close to I-610 north loop, Main Street & Sunset Boulevard between Rice University and Hermann Park, and South Rice Avenue & Elm Street off the I-610 west loop.

The global colocation pattern between no injury crashes and TL intersections (GCLQ = 7.90, significant at the 0.001 level) is also found unstable from place to place. In Figure 2C, the local analysis identified 51 no injury crashes that were significantly affected by TL intersections. Most of these incidents were also clustered in downtown Houston.

We also calculated LCLQs with SS intersections across three types of crashes. As shown in Figure 2D, five fatality crashes were significantly colocated with SS intersections. Of these, four crashes fall within the I-610 loop. Figure 2E illustrates the LCLQs of injury crashes with SS intersections. There were 109 injury crashes significantly attracted to SS intersection locations and were mainly distributed to the east of Houston within the I-610 loop. In Figure 2F, we saw 118 no injury crashes that were spatially correlated with SS intersections, and they were more inclined towards east Houston.

Figure 2G describes the LCLQs of fatality crashes with NC intersections. The global analysis reported significant dispersion pattern (GCLQ = 0.5, at the 0.001 level). The local analysis identified two fatality crashes significantly colocated with these intersections (average LCLQs = 1.29), however. In Figure 2H, we found 23 injury incidents across the city that were significantly attracted to NC intersections. And Figure 2I shows 39 no injury crashes scattered beyond the I-610 loop that were affected by NC intersections.

The above results indicate that the global patterns with any category of intersections across three types of crashes do not hold from place to place. The local analysis goes a step further than the global by showing how the global patterns were only valid at certain spots. Such drilling down from the global to local is essential because of the heterogeneous and quickly changing land use patterns and built environment. While a global measure helps identify patterns across the city, the local measure points out differences between various urban spaces and identifies the locations that need special attention.

Even though the variability of colocation (or dispersion) pattern with a certain type of intersections is not substantial across injury severity groups (especially for SS and NS intersections) by the global analysis in Table 1, we found notable disparities in their local colocation patterns. In terms of crashes that were significantly colocated with TL intersections, an evident cluster in downtown Houston was highlighted for fatality crashes, a scattering pattern mostly over the I-610 loop was identified for injury crashes, and the same dispersion layout further spread out the entire city was diagnosed for no injury crashes. The varying spatial distribution patterns help policy-makers identify troublesome areas more efficiently.

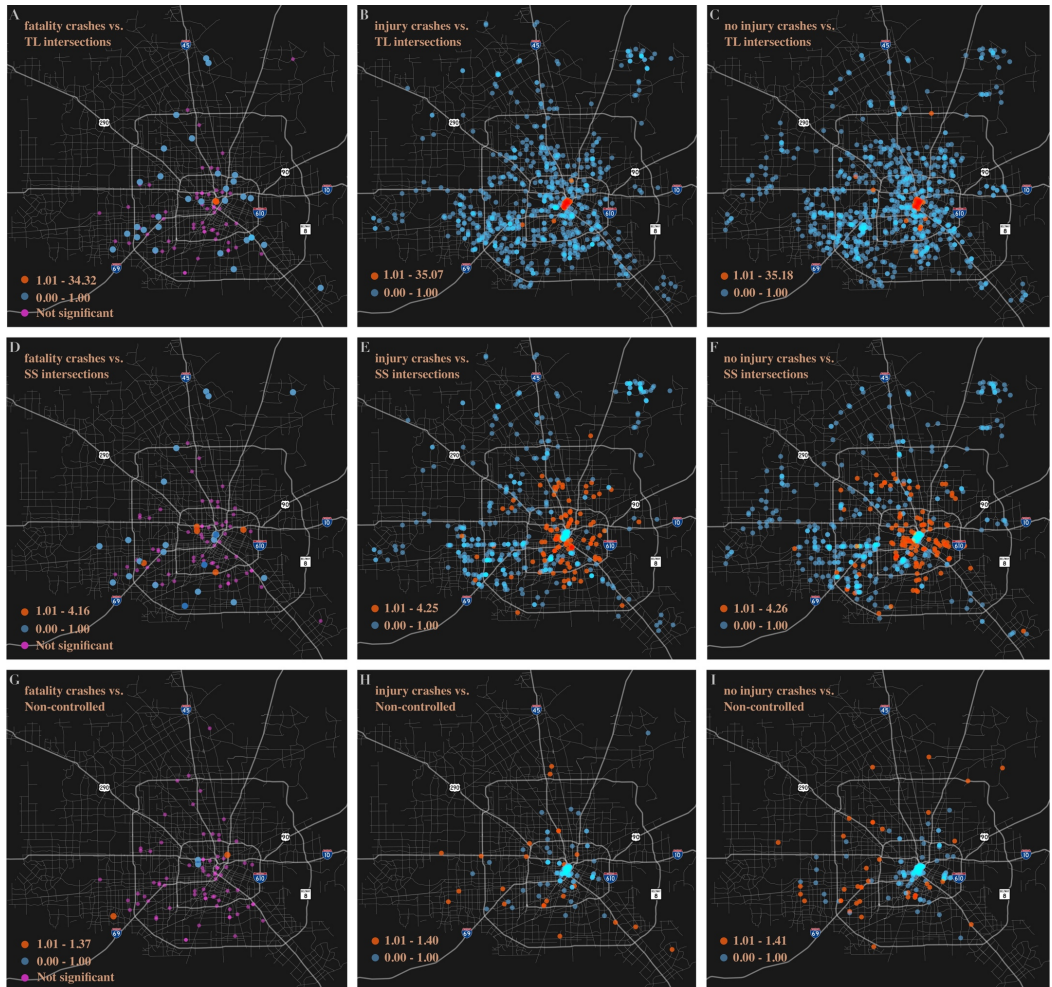

Figure 2. LCLQs of crashes and intersections: (A) fatality, (B) injury, and (C) no injury crashes vs. TL intersections; (D) fatality, (E) injury, and (F) no injury crashes vs. SS intersections; (G) fatality, (H) injury, and (I) no injury crashes vs. NC intersections

## 6. Discussions of Attributes to High Crash Risks

The global measure helps identify particular physical characteristics of potentially problematic intersections across the city (e.g., TL intersections), while the local highlights some specific locations of these types of intersections (e.g., Rusk Street & Louisiana Street). In this section, we will further investigate possible factors leading to high crash risks at these highlighted spots in Houston Area. We specifically focus on TL intersections here given their links with severe crash risks.

### 6.1 TL intersections

Figure 3 puts the 101 crashes across all three injury types that were significantly colocated with TL intersections together (injury severity type is distinguished by colors). Clearly, downtown remained the focal point of incidents. Beyond this area, three significantly linked injury crashes were identified in the I-610 loop and eight no injury crashes were observed—four incidents close to the I-610 south loop and four incidents between I-610 loop and Beltway 8 (the second loop in Houston). To more fully examine these crash hotspots and identify possible underlying factors, the 65 TL intersections related to these crash hotspots are drawn out in Figure 4. Looking closely at the structure of these highlighted TL intersections can help us identify possible reasons why crash incidents are the most likely to occur next to them and to begin thinking about ways to make these intersections safer for all users.

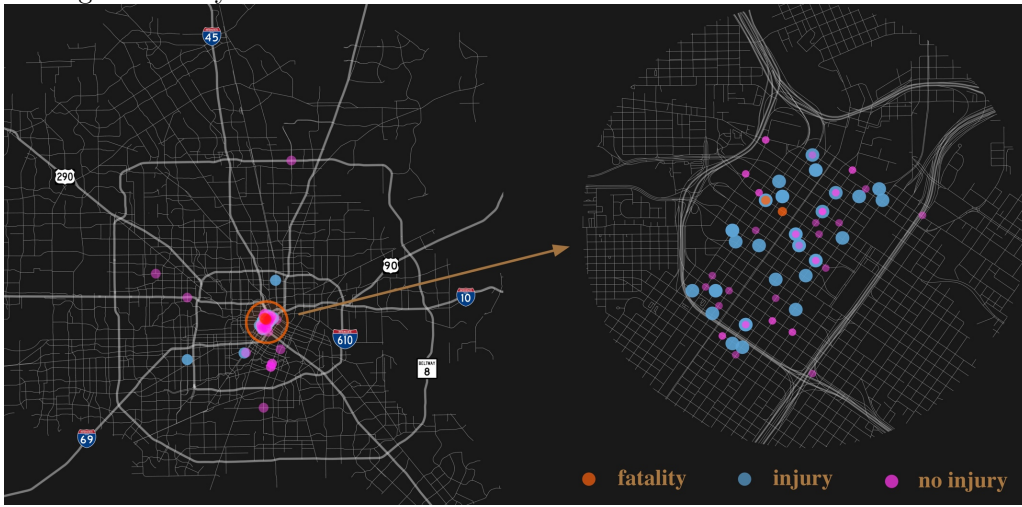

Figure 3. Locations of three types of crashes significantly colocated with TL intersections

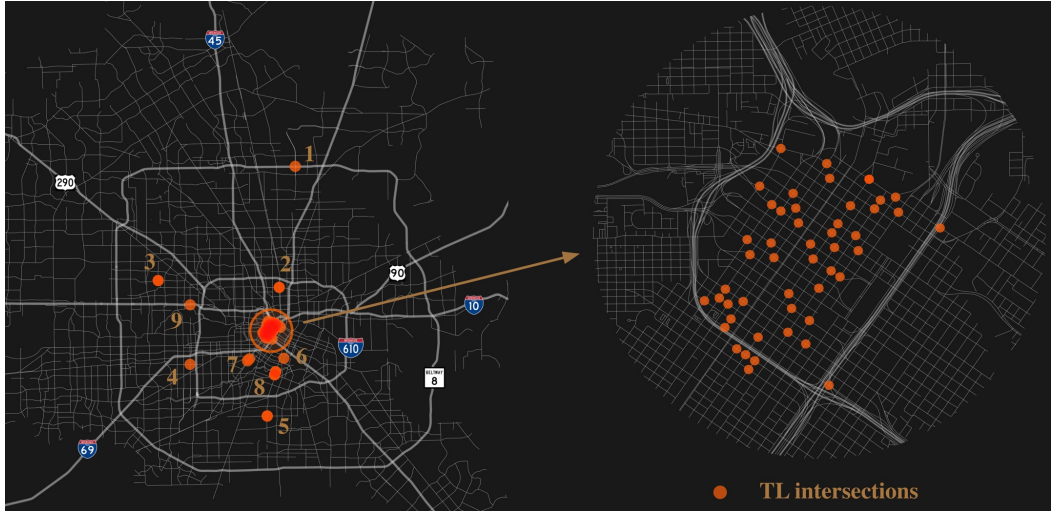

Figure 4. Locations of TL intersections that significantly attracted crashes across three injury types

As shown in Figure 4, the majority of the 65 TL intersections were clustered in downtown Houston. What makes the TL intersections in downtown more prone to crashes than those in other areas? First, downtown area has a high mixture of transportation modes. Pedestrians, bicycles, light rail, buses, and private vehicles all share the same spaces; and vehicles, pedestrians, and cyclists can cross the path of at-grade light rail with few barriers. What makes it worse is that pedestrians, vehicles, and light rail all have different control signals. The signals are timed differently resulting in possible confusion of which signals to follow. Second, the density of TL intersections is extremely high in downtown. There are multiple signals in very short distance, which may distract drivers and make them look at the wrong signal further down the street. Third, many one-way streets in downtown make pedestrians and cyclists face confused drivers, especially visitors who are unfamiliar with this setup. Fourth, Houston has dense bus routes in downtown area and many bus stops are set close to intersections. This layout brings additional activities to the already busy intersections. Finally, downtown is in a constant state of redevelopment with nearly constant construction. The on-going construction leads to sidewalk and street closures and causes additional confusion for drivers and therefore puts pedestrians and cyclists into more vulnerable positions.

Given the noticeable clusters of the high-risk intersections in downtown Houston, greater attention should be paid to this area. Engineers and policy makers may consider improving the coordination of the various types of traffic control signals, consistency of construction, safe traffic speeds, and caution signage and barriers around at-grade rail at or around intersections in downtown for safety improvement. Shifting bus stops away from intersection near-side corners to far-side or mid-block locations would likely reduce risks as well (Quistberg et al. 2015; Retting et al. 2003).

Clearly, there are a few other TL intersections with high crash risks beyond downtown Houston. For example, intersection 1 in Figure 4 is located at John F. Kennedy Boulevard and Texas 8 Beltway Frontage Road, the southern border of George Bush Intercontinental Airport. Intersection 2 is at Hardy Street and Cavalcade Street. Intersection 3 sits at Hammerly Boulevard and Hollister Road. Intersection 4 is located at South Rice Avenue and Gulfton Street. Intersection 5 is where Reed Road and Scott Street converge. Intersections 6, 7, 8 and 9 are related to University of Houston campus, Rice University campus—Hermann Park area, Brays Bayou Greenway Trail, and IKEA home furnishings company, respectively. Several common factors across these intersections are identified, for example, high posted speed limits, wide intersections, a mixture of modes, numerous traffic control mechanisms, and corner bus stops. Additional relevant factors include close proximity to schools and college campuses (Cunningham Elementary School at intersection 4, Worthing High School at intersection 5, University of Houston campus at intersection 6, and Rice University campus at intersection 7). That areas close to educational facilities are correlated with more pedestrian crashes is also noted by Loukaitou-Sideris et al. (2007).

**6.2 SS intersections and NC intersections**

The local analysis identified 232 crash incidents that were significantly tied to SS intersections across all three injury types with a mean LCLQ of 4.2. It also identified 64 crashes regardless of injury types significantly attracted to NC intersections with a mean LCLQ of 1.3. The mean LCLQs between crashes and SS and NC intersections are far less pronounced than that between all three crash types and TL intersections, however. This suggests that SS and NC intersections are far less likely to attract crashes than TL intersections. In Figure 5, we map the locations of the related 232 SS intersections and 55 NC intersections. It is easy to see that SS intersections were mostly located in central Houston (within I-610 loop) while NC intersections were scattered outside the loop to the west and often along major roadways.

Visually examining the intersections in Figure 5 leads to some general observations. Most of these SS intersections are at junctures connecting local streets to major roads and are found relevant to high crash risks (Laberge et al. 2006). These major roads are often wide streets with a high posted speed limit and that have bus stops that pedestrians are likely to access. Take Hullsmith Drive & Westheimer Road highlighted in Figure 5 as an example. Westheimer Road is a wide (eight-lane) street with busy traffic. There is a near-side bus stop located on the other side of this intersection and it is the only eastbound stop on the popular 82 bus route. To access it from the north users must either cross eight lanes of traffic without aid (no crosswalks) or walk to the closest TL intersection and then walk back. But it would be difficult for them to do so as there is no signal for pedestrians crossing the intersection. The closet TL intersection with crosswalks and pedestrian signals is 0.3 miles away, a distance greater than most people are likely to walk even under ideal conditions. Possible countermeasures to improve the traffic safety of these SS intersections could be to install additional pedestrian safety controls—mid-block signalized crosswalks, or more traffic lights in strategic areas.

The NC intersections in Figure 5 are usually the egress of shopping centers, apartment gates or personal houses that connects to major roads and are deemed to be most hazardous for pedestrian crashes because there are no crosswalks, lights, signs, or other indicators installed at these intersections to alert users of the street (Ossenbruggen et al. 2001). In addition, many of these intersections are found leading onto relatively high-trafficked streets with higher speed traffic than a typical residential street would have. As shown in Figure 5, Park Row Drive near Park Falls Community is one such example. The egress from the community has no stop sign or other traffic control signal. Drivers are expected to yield to crossing traffic, but there is no indication of proper pedestrian spaces or formal regulation of driver behavior. We suggest putting noticeable pedestrian cross signs and adding crosswalks to help improve the visibility of pedestrian space and raise the awareness of drivers of possible pedestrians and hence lower the crash risk.

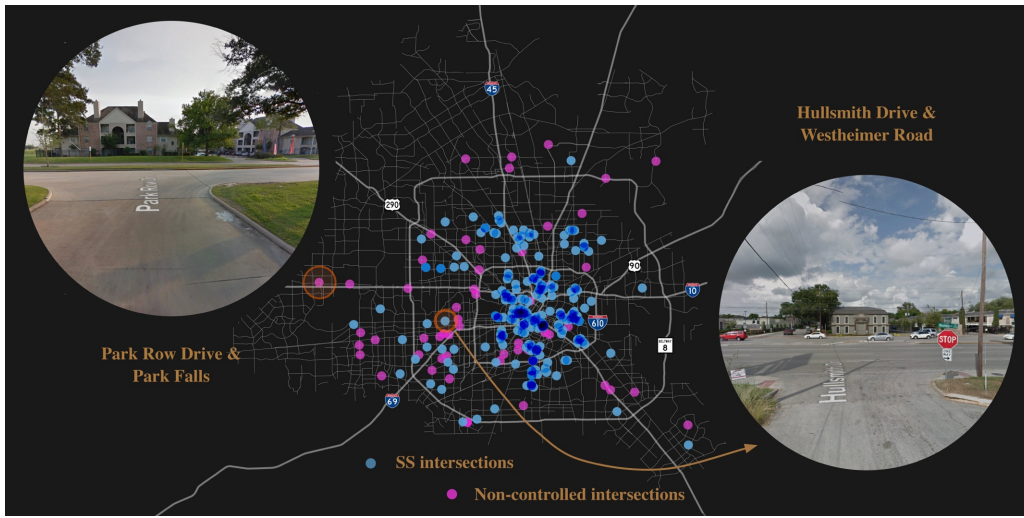

Figure 5. Locations of SS and NC intersections that significantly attracted crashes across three injury types

## 7. Conclusions

Given the increasing number of crashes and injuries to pedestrians and cyclists, stakeholders across the nation—from cities, to law enforcement agencies, to citizen advocates—are working on ways to improve traffic safety. The need for this focus will only grow as our cities and suburbs densify and walking/biking becomes more popular. Although many studies have targeted this issue by, for instance, building regression models to understand the ways elements of the built environment are connected to crash rates in an area, their work may not provide policy-makers with a detailed look at crash patterns. This research proposes a novel way to understand at which intersections crashes are more likely to occur. We developed two colocation indicators, the GCLQ and LCLQ, to examine the spatial correlation patterns between crashes and intersections. Specifically, the GCLQ evaluates the overall associations in an entire region, while the LCLQ assesses individual patterns at each crash location. Compared to traditional macro-level models, the LCLQ analysis is able to extract site-specific patterns and identify, among so many controlled intersections, the ones that are most dangerous to pedestrians and cyclists. Our analysis might be the first effort to demonstrate the viability of studying spatial correlation patterns to identify dangerous intersections. In addition, the LCLQ method is free from the well-known MAUP effect as a result of its disaggregate design and hence can provide more reliable results. Such features can facilitate stakeholders to make more informed and reliable decisions about future street-level safety improvements.

One limitation of this study is that the data, including the city boundary, road network, locations of traffic lights and stop signs, for each year between 2010 and 2016 is not completely consistent. This paper focuses on the methodological end, and we hope to be able to gather more consistent data and report our future work.

## Acknowledgements

Hu would like to acknowledge the support from the Houston Endowment. We thank the editor and the three anonymous referees for their valuable comments that greatly improved the manuscript.